\documentstyle[preprint,aps]{revtex}

\def\8{\infty}

\def\undertext#1{\vtop{\hbox{#1}\kern 1pt \hrule}}

\def\VEV#1{\left\langle\,#1\,\right\rangle}

\def\br{\\ \nonumber & &}

\def\be{\begin{equation}}
\def\ee{\end{equation}}
\def\bea{\begin{eqnarray} & &}
\def\eea{\end{eqnarray}}
\def\ct#1{\cite{#1}}
\def\rf#1{(\ref{#1})}

\def\t{\tilde}

\begin{document}

\preprint{IASSNS-HEP-97/5, NSF-ITP-97-014}

\title{The Haldane-Rezayi Quantum Hall State and Conformal
Field Theory}

\author{V. Gurarie\footnote{Research supported by NSF grant DMS 9304580
         ~~~~gurarie@ias.edu}
         and M. Flohr\footnote{Research supported by the 
                      Deutsche Forschungsgemeinschaft  ~~~~flohr@ias.edu}}
 
\address{Institute for Advanced Study,
Olden Lane,
Princeton, N.J. 08540}

\author{C. Nayak\footnote{Research supported by NSF grant PHY94-07194
 ~~~~nayak@itp.ucsb.edu} }

\address{Institute for Theoretical Physics,
University of California,
Santa Barbara, CA, 93106-4030}

\date{\today}
\maketitle

\begin{abstract}
We propose field theories for the bulk and edge
of a quantum Hall state in the universality class
of the Haldane-Rezayi
wavefunction. The bulk theory is associated with the $c=-2$
conformal field theory. The topological properties
of the state, such as the quasiparticle braiding
statistics and ground state degeneracy on a torus,
may be deduced from this conformal field theory.
The 10-fold degeneracy on a torus is explained by the
existence of a logarithmic operator in the
$c=-2$ theory; this operator corresponds to a novel
bulk excitation in the quantum Hall state.
We argue that the edge theory is the $c=1$ chiral Dirac fermion,
which is related in a simple way to the
$c=-2$ theory of the bulk.
This theory is reformulated as a truncated version of a doublet
of Dirac fermions in which the $SU(2)$ symmetry -- which corresponds
to the spin-rotational symmetry of the
quantum Hall system -- is manifest and non-local.
We make predictions for the current-voltage
characteristics for transport through point contacts.

\end{abstract}
 
\pacs{PACS numbers:}
 
\narrowtext

\section{Introduction}

In 1987, Willet, {\it et al.} \cite{willet} discovered a fractional quantum
Hall plateau with conductance ${\sigma_{xy}} = \frac{5}{2}\,
\frac{e^2}{h}$. Shortly thereafter, Haldane and Rezayi \cite{hr}
suggested
\begin{equation}
{\Psi_{\rm HR}}\,\, =\,\,
{\cal A}\left(\frac{{u_1}{v_2}-{v_1}{u_2}}{({z_1} - {z_2})^2}\,\,\,
\frac{{u_3}{v_4}-{v_3}{u_4}}{({z_3} - {z_4})^2}\,\,\,\ldots\,\right)
\,\,\,{\prod_{i>j}}{\left({z_i} - {z_j}\right)^2}\,\,\,
{e^{-\frac{1}{4{\ell}_0^2} \sum |z_i|^2 }}
\label{hrstate}
\end{equation}
(${\cal A}$ means antisymmetrization over all exchanges of
electrons,  $u_i$, $v_i$ are respectively up and down spin
states of the $i^{th}$ electron, and $\ell_0$ is the magnetic length)
as a variational {\it ansatz} for the incompressible
state of electrons observed in this experiment.\footnote{(\ref{hrstate})
is a wavefunction for electrons at $\nu=1/2$. It is assumed that
at $\nu=5/2$ the lowest Landau levels of both spins are
filled and the appropriate analog of (\ref{hrstate})
in the second Landau level describes the additional $\nu=1/2$.}
Despite the passage of nearly 10 years, this
proposal has been neither confirmed nor ruled
out by experiment, in large part because the
theoretical understanding of this state is still primitive.
In this paper, we try to address this deficiency by
proposing effective field theories of the bulk and
edge of a system of electrons with ground state
given by (\ref{hrstate}).

Our faith in Laughlin's wavefunctions for the principal
odd-denominator states stems not only from their large
overlap with the exact ground state of small systems.
Rather, their success lies in the
fact that they exhibit properties -- `topological
ordering' \cite{wentoporder,landauginz} -- which are plausibly
far more robust than the specifics of any trial wavefunction.
The `topological ordering', which refers to the fractional statistics
of quasiparticles \cite{halpstat,fracstat}
and off-diagonal long-range-order
of certain non-local order parameters \cite{landauginz},
could be demonstrated for Laughlin's wavefunctions
because the plasma analogy facilitates calculations
with these wavefunctions. The `topological ordering'
is summarized by the (abelian) Chern-Simons effective
field theories of the quantum Hall effect
(see also \cite{capelli,christofano,frohlich}).
The fractional statistics is the linchpin of the
theory and its most startling prediction,
and, hopefully, will be confirmed some day soon.
The Chern-Simons effective field theory led, in turn,
to a conformal field theoretic description of the edge
excitations \cite{wenedge}. Detailed predictions based on the edge theory
have recently been spectacularly confirmed
\cite{kanefisher,fendley,webb,chang}.
Unfortunately, there is no plasma analogy for
the Haldane-Rezayi wavefunction, nor for a number
of other proposed wavefunctions such as the Pfaffian
state. Consequently, the correct Chern-Simons theory
of the quasiparticle statistics and the conformal
field theory of the edge excitations have remained elusive.

A way of skirting this obstacle was suggested by
Moore and Read \cite{mooreread}. They observed that the Laughlin state
and a number of other quantum Hall effect wavefunctions,
including the Haldane-Rezayi state, could be interpreted as
the conformal blocks of certain conformal field theories.
This observation gains great power in light
of the equivalence, discovered by Witten \cite{witten},
between the states of a Chern-Simons theory and
conformal blocks in an associated conformal field theory.
It is often the case that a quantum Hall state can be
reproduced by the conformal blocks of a
conformal field theory. Since this
conformal field theory is equivalent to
a Chern-Simons theory, it is very tempting to close the circle,
following Moore and Read, and conjecture that this
Chern-Simons theory is the desired effective theory of
the bulk, which
would be obtained by a direct calculation using
brute force or some generalization of the
plasma analogy. This conjecture is true for
states with abelian statistics, such as the Laughlin
states and their hierarchical descendents.
A general argument in favor of this
premise was given in \cite{wilczek}, where it was used to deduce
the $SO(2n)$ non-abelian statistics of quasiholes
in the Pfaffian state once the correspondence
between this state and the conformal blocks of the
$c = \frac{1}{2} + 1$ conformal field theory was demonstrated
in some detail. If correct, this conjecture implies
that the conformal blocks are the preferred basis
of the quantum Hall wavefunctions since they make
the quasiparticle statistics transparent.

It has been observed \cite{mooreread,wenwu,milovan}
that the Haldane-Rezayi
state is a conformal block in the $c=-2$ conformal
field theory. Here, we derive some consequences of this fact.
Among these is the 10-fold degeneracy of the ground
state on the torus. The ground state degeneracy on
a torus is not merely
mathematical trivia. It is
equal to the number of `topologically dictinct'
quasihole excitations -- ie. that have inequivalent
braiding properties (so, for instance, the combination
of any excitation with a bosonic excitation does
not produce a topologically distinct excitation)
-- which there are in the system. As we will explain, the
10-fold degeneracy is a surpise, and is due to the existence
of an excitation which cannot be found in other
proposed even-denominator quantum Hall states, such
as the Pfaffian and $(3,3,1)$ states. The 10-fold
degeneracy is due, in the $c=-2$ conformal field theory,
to the existence of a logarithmic operator \cite{Gur}. We
elucidate the structure of the $c=-2$ theory,
with particular emphasis on the calculation of conformal
blocks and on the logarithmic operator. The former
allow us to obtain the non-abelian statistics
of the quasiholes.

In principle, it should be possible to use the Chern-Simons
theory of the bulk to deduce the conformal field theory
of the edge excitations. However, a more direct approach
is possible. As can be explicitly shown for the Laughlin states
\cite{symmpol} (see, also, the second ref. in \cite{wenedge}),
the states of the edge conformal field theory can be
enumerated by direct construction of the corresponding
lowest Landau level wavefunctions which are the exact
zero-energy eigenstates of certain model Hamiltonians.
Under mild asumptions about the confining potential at
the edge of the system, which gives these excitations
non-zero energy, the energy spectrum can be obtained as well.
Milovanovic and Read \cite{milovan} generalized this construction
to the Pfaffian, $(3,3,1)$, and Haldane-Rezayi states.
In the case of the Pfaffian and $(3,3,1)$ states, their
construction led immediately to the correct edge theory.
We propose that the edge theory of the Haldane-Rezayi
state is a theory of a chiral Dirac fermion, with
$c=1$. This theory possesses a global $SU(2)$ symmetry
which becomes manifest when recast as
a truncated version of a $c=2$ theory. The $SU(2)$
symmetry -- which is just the spin-rotational symmetry,
an approximate symmetry in GaAs with its small
$g$ factor and effective mass -- is unusual in that the
local spin-densities do not have local commutation
relations (see, also, \cite{ludwig}). This indicates the impossibility of
localizing spin at the edge which, we argue, is
supported by an analysis of the explicit wavefunctions.
The relationship between the $c=-2$ and $c=1$ theories 
\cite{Kausch,Flohr,ludwig} --
they have nearly the same states, spectra, and, therefore,
partition functions -- is very natural in this context
since these theories describe the bulk and edge of the
Haldane-Rezayi state.

Section II is a review of the relevant facts and standard lore
regarding the Haldane-Rezayi state. In section III we
recapitulate, in order to make our exposition as self-contained
as possible, the conformal field theory approach
to the bulk wavefunctions in the quantum Hall effect.
In section IV, we discuss the $c=-2$ conformal field theory
and, in section V, we apply it to study quasiparticles
in the Haldane-Rezayi state. Section VI is devoted to
the relationship between the $c=-2$ and $c=1$ conformal
field theories. The edge theory of the Haldane-Rezayi state
is discussed in section VII and the physical consequences
following from the results of sections V and VII are
discussed in section VIII. Parts of this paper are rather technical.
Readers who are uninterested in the subtleties and finer points of the
$c=-2$ and $c=1$ conformal field theories but are interested
in the observable consequences which follow
from them may wish to skip or merely skim
sections III, IV, and VI.

\section{The Haldane-Rezayi State}

In this paper, we will be studying the zero-energy
eigenstates of the Hamiltonian \cite{hr}
\be
H = {V_1}\,\,{\sum_{i>j}}\, \delta'({z_i}-{z_j}),
\label{halrezham}
\ee
where ${V_1}>0$.
While this is almost certainly not the Hamiltonian
governing any experiment, it has the advantage of tractability,
and the properties which interest us are universal and
should be stable against perturbations. We have assumed that
the Zeeman energy vanishes so (\ref{halrezham}) is
invariant under $SU(2)$ spin rotations. In GaAs,
the Zeeman energy is $\frac{1}{60}$ of the cyclotron energy,
so $SU(2)$ will be a reasonable approximate symmetry.

This Hamiltonian shares with other simple, short-range
lowest Landau level Hamiltonians the property that
not only the ground state, but also states with quasiholes
and edge excitations, have zero energy. This should not be
troubling since the incompressibility of the quantum Hall
state depends upon the existence of a discontinuity in the
chemical potential. If quasiparticles have finite
energy but quasiholes do not, there will be such a discontinuity.
For a more realistic interaction,
both quasiholes and quasiparticles have finite energy, but
the discontinuity persists.
Since they are annihilated by the Hamiltonian, the
multi-quasihole states are easier to enumerate, so in
what follows, we will discuss them exclusively; the properties
of quasiparticles -- though difficult to study directly --
are trivially related to those of quasiholes.
The vanishing energy of the edge excitations should
not be a surprise, either. These have finite energy
only if there is a confining potential at the edge of
the system, as there will be in any real 2D electron gas.
As we discuss further below, we will simply assume that,
in the presence of a confining potential,
the energy of an edge excitation is proportional to its
angular momentum.
For these reasons,
we will refer to the state with no quasiholes and no
edge excitations -- the maximally compressed state --
as `the ground state' and refer to the other
zero-energy states as `quasihole states' and `edge states',
respectively, despite the fact that, strictly speaking,
all of these states are ground states of (\ref{halrezham}).
The multi-quasihole states, which are bulk excitations,
can be distinguished from the edge states by the
property that the former must be homogeneous in the $z_i$'s since
only such wavefunctions can be extended to the sphere
(where there is no edge and, hence, no edge excitations).
The inhomogeneous zero-energy wavefunctions are
edge states.

As we mentioned above, the ground state of (\ref{halrezham})
is the Haldane-Rezayi state,
\be
{\Psi_{\rm HR}}\,\, =\,\,
{\rm Pf}\left(\frac{{u_i}{v_j}-{v_i}{u_j}}{({z_i} - {z_j})^2}\right)
\,\,\,{\prod_{i>j}}{\left({z_i} - {z_j}\right)^2}\,\,\,
{e^{-\frac{1}{4{\ell}_0^2} \sum |z_i|^2 }}
\label{hrstateb}
\ee
where ${\rm Pf}$, the Pfaffian, is the square root of the
determinant of an anti-symmmetric matrix, or, equivalently,
the antisymmetrized product over pairs introduced in
(\ref{hrstate}). It resembles its cousins, the `Pfaffian'
and $(3,3,1)$ states:
\be
{\Psi_{\rm Pf}}\,\, =\,\,
{\rm Pf}\left(\frac{{u_i}{u_j}}{{z_i} - {z_j}}\right)
\,\,\,{\prod_{i>j}}{\left({z_i} - {z_j}\right)^2}\,\,\,
{e^{-\frac{1}{4{\ell}_0^2} \sum |z_i|^2 }}
\label{pfstate}
\ee
\be
{\Psi_{\rm (3,3,1)}}\,\, =\,\,
{\rm Pf}\left(\frac{{u_i}{v_j}+{v_i}{u_j}}{{z_i} - {z_j}}\right)
\,\,\,{\prod_{i>j}}{\left({z_i} - {z_j}\right)^2}\,\,\,
{e^{-\frac{1}{4{\ell}_0^2} \sum |z_i|^2 }}
\label{ttostate}
\ee
The Pfaffian factors are reminiscent of the real space
form of the BCS pairing wavefunction. The Haldane-Rezayi
state can be interpreted as a quantum Hall state of
spin-singlet $d$-wave pairs while the Pfaffian and
$(3,3,1)$ states can be interpreted as spin-triplet
$p$-wave paired states with ${S_z} = 1$ and ${S_z} = 0$,
respectively. These states are discussed in
\cite{halphelv,mooreread,greiter,ho,wilczek,readrezayib}.

The quasiholes in the Haldane-Rezayi state are,
like the vortices in a superconductor, half
flux quantum excitations. A wavefunction for a state with
two such quasiholes at $\eta_1$ and $\eta_2$
can be written down by modifying the factor
inside the Pfaffian in (\ref{hrstateb}):
\be
\Psi\,\, =\,\,
{\rm Pf}\left(\frac{\left({z_i} - {\eta_1}\right)
\left({z_j} - {\eta_2}\right) + i\leftrightarrow j}
{({z_i} - {z_j})^2}\,\, \left({u_i}{v_j}-{v_i}{u_j}\right)\right)
\,\,\,\,{\prod_{i>j}}{\left({z_i} - {z_j}\right)^2}
\label{hrtqhstate}
\ee
Here, and henceforth, we will be sloppy and omit
the Gaussian factor in the wavefunction so as to avoid
excessive clutter.
The half flux quantum quasiholes have charge $\frac{1}{4}$.
As in the case of the Pfaffian and $(3,3,1)$ states,
there is not a unique state of $2n$ quasiholes
at ${\eta_1},{\eta_2},\ldots,{\eta_{2n}}$. Rather,
there is a degenerate set of states. This degeneracy
is the {\it sine qua non} for non-Abelian statistics.
Consider the four-quasihole case. Define the three
polynomials
\be
{P_\sigma}({z_i},{z_j})
= \left({z_i} - {\eta_{\sigma(1)}} \right)\,
\left({z_i} - {\eta_{\sigma(2)}} \right)\,\,
\left({z_j} - {\eta_{\sigma(3)}} \right)\,
\left({z_j} - {\eta_{\sigma(4)}} \right)\,\,+\,\,
i\leftrightarrow j
\ee
where $\sigma$ is a permutation of $\{1,2,3,4\}$.
The following four-quasihole states
\be
\Psi\,\, =\,\,
{\rm Pf}\left(\frac{{P_\sigma}({z_i},{z_j})}
{({z_i} - {z_j})^2}\,\, \left({u_i}{v_j}-{v_i}{u_j}\right)\right)
\,\,\,\,{\prod_{i>j}}{\left({z_i} - {z_j}\right)^2}
\label{hrmqhstate}
\ee
are annihilated by (\ref{halrezham}). These wavefunctions
are not all linearly independent. Linear relations,
found in \cite{wilczek}, reduce the set (\ref{hrmqhstate})
to a basis set of $2$ linearly independent states.
There are also states which are not spin-singlets.
When there are $2n$ quasiholes, there are $2^{2n-3}$ 
linearly independent states; the following particularly
elegant basis was found in \cite{readrezayib}:
\bea
\label{rrmqhwfcn}
\Psi\,\, =\,\,
{\cal A}\biggl({z_1^{p_1}}{\chi_1}\ldots{z_{k-1}^{p_{k-1}}}{\chi_{k-1}}
\frac{\left({u_k}{v_{k+1}}-{v_k}{u_{k+1}}\right)\,\,
{P_\sigma^{2n}}({z_k},{z_{k+1}}) }
{({z_k} - {z_{k+1}})^2}\br\times\,\,
\frac{\left({u_{k+2}}{v_{k+3}}-{v_{k+2}}{u_{k+3}}\right)\,\,
{P_\sigma^{2n}}({z_{k+2}},{z_{k+3}})}
{({z_{k+2}} - {z_{k+3}})^2}\,\,\,\ldots\,\biggr)
\,\,\,{\prod_{i>j}}{\left({z_i} - {z_j}\right)^2}
\eea
where $\chi_i$ is the spin wavefunction of the $i^{th}$
electron, $k\leq n$, ${p_j}\leq n-2$ and $\sigma$ is some permutation
which is fixed once and for all. The most general
multi-quasihole excitation can also have charge $\frac{1}{2}$
`Laughlin quasiholes' at
${\lambda_1}, {\lambda_2},\ldots,{\lambda_l}$,
\bea
\label{hrgmqhstate}
\Psi\,\, =\,\,
{\cal A}\biggl({z_1^{p_1}}{\chi_1}\ldots{z_{k-1}^{p_{k-1}}}{\chi_{k-1}}
\frac{\left({u_k}{v_{k+1}}-{v_k}{u_{k+1}}\right)\,\,
{P_\sigma^{2n}}({z_k},{z_{k+1}}) }
{({z_k} - {z_{k+1}})^2}\br\times\,\,
\frac{\left({u_{k+2}}{v_{k+3}}-{v_{k+2}}{u_{k+3}}\right)\,\,
{P_\sigma^{2n}}({z_{k+2}},{z_{k+3}})}
{({z_{k+2}} - {z_{k+3}})^2}\,\,\,\ldots\,\biggr)
{\prod_{i,\alpha}}\left({z_i} - {\lambda_\alpha}\right)
\,\,\,{\prod_{i>j}}{\left({z_i} - {z_j}\right)^2}
\eea
Although the charge $\frac{1}{2}$ `Laughlin quasiholes'
can be made by bringing together two charge $\frac{1}{4}$
quasiholes, we distinguish them because they do not
affect the Pfaffian, or `pairing,' part of the wavefunction.

It is instructive to consider the Haldane-Rezayi state on a
torus. The Hamiltonian (\ref{halrezham}) no longer has
a unique ground state. Its degenerate ground states
are:
\be
{\Psi_{\rm HR}^{a,b}}\,\, =\,\,
{\rm Pf}\left(
\frac{\left({u_i}{v_j}-{v_i}{u_j}\right)\,\,\,
{\vartheta_a}({z_i} - {z_j}){\vartheta_b}({z_i} - {z_j})}
{{\vartheta_1^2}({z_i} - {z_j})}
\right)
\,\,\,{\prod_{i>j}}{\vartheta_1^2}\left({z_i} - {z_j}\right)\,\,\,
{\prod_{k=1}^2}{\vartheta_1}\left({\sum_i}{z_i}-{\zeta_k}\right)
\label{hrstatetorus}
\ee
where $\zeta_k$, $k=1,2$ are arbitrary complex numbers. Here
$a,b = 2,3,4$, but there is a linear relationship between
${\vartheta_2^2},{\vartheta_3^2},{\vartheta_4^2}$, so there is a
5-fold degeneracy arising from this choice\footnote{For the definition of
the standard elliptic $\vartheta$-functions and their use in constructing
the wave functions on a torus see, for example, ref. \ct{readrezayib}.}. 
There is an
additional factor of two from the choice of the
$\zeta_k$'s, so the total ground state degeneracy
is 10 (see, also, \cite{vakkuri}).
The degeneracy on a torus is not only an important way
of distinguishing quantum Hall states found in numerical studies,
but also has a simple physical significance. The different
degenerate ground states are obtained from each other
by creating a quasihole-quasiparticle pair, taking one
around a non-trivial cycle of the torus, and annihilating
them. There are as many degenerate ground states as
there are distinct, non-trivial ways of doing this.
In other words, the ground state degeneracy on a torus
is equal to the number of {\it distinct}
bulk excitations that the quantum Hall state admits,
where {\it distinct} refers to the braiding properties
of the excitations. We will return to this issue
in the next section.

As in the case of the bulk excitations, the edge
excitations may be naturally divided into a direct
product of those which do not affect the pairing part
of the wavefunction with those which only affect
the Pfaffian factor. The former are generated
by multiplying the ground state by symmetric
polynomials. They form a $1+1$-dimensional
chiral bosonic theory with $c=1$. In a Laughlin
state at $\nu=\frac{1}{m}$, these would be sufficient
to span the space of edge excitations. In the
Haldane-Rezayi state, however, we also have
the wavefunctions which modify the pairing
part of the wavefunction \cite{milovan}. These are closely
related to the  form (\ref{rrmqhwfcn}) of the
multi-quasihole wavefunctions
\begin{equation}
\Psi\,\, =\,\,
{\cal A}\left({z_1^{p_1}}{\chi_1}\ldots{z_{k-1}^{p_{k-1}}}{\chi_{k-1}}
\frac{{u_k}{v_{k+1}}-{v_k}{u_{k+1}}}
{({z_k} - {z_{k+1}})^2}\,\,\,\,
\frac{{u_{k+2}}{v_{k+3}}-{v_{k+2}}{u_{k+3}}}
{({z_{k+2}} - {z_{k+3}})^2}\,\,\,\ldots\,\right)
\,\,\,{\prod_{i>j}}{\left({z_i} - {z_j}\right)^2}
\end{equation}
In these states, $k-1$ of the electrons are
unpaired. There is no restriction on the $p_j$'s, so
the unpaired electrons increase the angular momentum
(the total number of powers of $z$)
above that of the ground state by ${p_1}+1,
{p_2}+1,\ldots,{p_{k-1}}+1$, with positive ${p_i}$,
no more than two of which may be equal
(because of the requirements of analyticity and
antisymmetry).
These electrons have spins ${\chi_1},\ldots,{\chi_{k-1}}$.
Dividing the $p_i$'s into those associated with
up-spin electrons, $n_i$'s, and those associated with
down-spin electrons, $m_i$'s,
this sector of the edge theory is composed of
states
\be
|{n_1},{n_2},\ldots,{n_\alpha};
\,{m_1},{m_2},\ldots,{m_\beta}\rangle
\label{hres}
\ee
with ${n_i}\neq{n_j}$, ${m_i}\neq{m_j}$ if
$i\neq j$. These states correspond to wavefunctions
\bea
\Psi\,\, =\,\,
{\cal A}\Biggl({z_1^{n_1}}{u_1}\ldots{z_{\alpha}^{n_{\alpha}}}{u_{\alpha}}
\,\,\,{z_{\alpha+1}^{m_1}}{v_{\alpha+1}}\ldots
{z_{\alpha+\beta}^{m_{\beta}}}{v_{\alpha+\beta}}\br
\frac{{u_{\alpha+\beta}}{v_{\alpha+\beta+1}}-
{v_{\alpha+\beta}}{u_{\alpha+\beta+1}}}
{({z_{\alpha+\beta}} - {z_{\alpha+\beta+1}})^2}\,\,\,\,
\frac{{u_{\alpha+\beta+2}}{v_{\alpha+\beta+3}}-
{v_{\alpha+\beta+2}}{u_{\alpha+\beta+3}}}
{({z_{\alpha+\beta+2}} - {z_{\alpha+\beta+3}})^2}\,\,\,\ldots\,\Biggr)
\,\,\,{\prod_{i>j}}{\left({z_i} - {z_j}\right)^2}
\label{eswf}
\eea
and have angular momenta (which, in the $1+1$-dimensional
edge theory, are just the momenta along the edge)
\be
{\sum_i}\left({n_i}+1\right)\,\,\,+
\,\,\,{\sum_i}\left({m_i}+1\right)
\ee
and
\be
{S_z} = \frac{1}{2}\left(\alpha - \beta\right)
\ee
This is simply a Fock space of spin-$\frac{1}{2}$
fermions. The fermions are neutral, since the
number of filled fermionic levels
can be changed without changing the electron
number.\footnote{This is not quite true. The fermion
number is congruent to the electron number
modulo 2; see \cite{milovan} for a dicussion of this
point. However, this does not affect the conclusion
that the fermionic excitations are neutral since
pairs of fermions can be created without changing
the charge.}
If we assume that the energy of a state is proportional
to its angular momentum relative to that of
the ground state (in general, it will be some
function of the angular momentum, which we expand in
powers of the angular momentum; the higher powers
will be irrelevant, in a renormalization group
sense), then the low-energy effective field theory
of the edge must be a theory of spin-$\frac{1}{2}$
neutral (and, hence, real) fermions. Since the
spin-$\frac{1}{2}$ representation of $SU(2)$ is
not a real representation, it is difficult to see
what this theory should be.  We return to this
puzzle in section VII.

The states (\ref{eswf}) are the vacuum sector of the
edge theory. There are also sectors in which charge
has been added to the edge. The wavefunctions for these
sectors have fractionally-charged quasiholes in the interior
which results in fractional charges being added to the edge.
When an integer charge is added to the edge, the
vacuum sector is recovered again. When a charge
corresponding to half-integral flux is added to the edge,
a quasihole of the type (\ref{hrtqhstate}) is present in 
the bulk. The edge excitations are still of the
form (\ref{eswf}), but the angular momenta associated
with them are now half-integral, ${n_i}+\frac{1}{2}$
and ${m_i}+\frac{1}{2}$. This is a `twisted sector'
\cite{milovan}.

\section{CFT for the Bulk of a Quantum Hall State}

The signature of a quantum Hall state is the braiding statistics
of the localized excitations, the quasiholes and quasiparticles.
Following \cite{fracstat}, we would calculate them using the Berry's phase
technique, according to which the states $|i\rangle$ transform as
\be
|i\rangle \rightarrow
{\rm P}exp\left(i\oint{\gamma_{ij}}\right)\,|j\rangle
\label{berryphase}
\ee
(${\rm P}exp$ is the path-ordered exponential integral)
when the $\alpha^{th}$ quasihole, with position $\eta_\alpha$,
is taken in a loop enclosing others, where
\be
{\gamma_{ij}} = \langle i|\frac{\partial}{\partial\eta_\alpha}|j\rangle
\label{berrymatel}
\ee
The fractional statistics of quasiholes in the Laughlin
states were established in this way \cite{fracstat}. However, the matrix
elements (\ref{berrymatel}) cannot be directly evaluated
for more complicated states such as the Haldane-Rezayi
state.

To calculate the braiding statistics of quasiholes in the
Haldane-Rezayi state, we will take the approach suggested
by Moore and Read \cite{mooreread}, which is, essentially, to guess
the Chern-Simons effective field theory of this state. To motivate
this guess, we look for a conformal field theory which has
conformal blocks which are equal to the quantum Hall
wavefunctions. As a warm-up, let's see how this works in the case
of the Laughlin state at $\nu=\frac{1}{m}$ where the Berry
phase calculation can be used as a check for the correctness of
this procedure.

The Hamiltonian
\be
H = {\sum_{k=0}^{m-1}}\,{V_k}{\sum_{i>j}} {\delta^{(k)}}({z_i}-{z_j})
\ee
annihilates the Laughlin ground and multi-quasihole (at positions
${\eta_1},\ldots,{\eta_n}$) states.
\be
{\Psi_{1/m}} = {\prod_{i>j}}\left({z_i}-{z_j}\right)^m
\label{laughgs}
\ee
\be
{\Psi_{1/m}^{\rm qh}} = {\prod_k}\left({z_k} - {\eta_1}\right)\,\,\ldots
{\prod_k}\left({z_k} - {\eta_n}\right)\,
{\prod_{i>j}}\left({z_i}-{z_j}\right)^m
\label{laughqhs}
\ee
The ground state
is equal to the following conformal block in the $c=1$ theory
of a chiral boson, $\phi$, with compactification radius
$R=\sqrt{m}$:
\be
{\Psi_{1/m}} = \langle\, {e^{i\sqrt{m}\phi({z_1})}}\,
{e^{i\sqrt{m}\phi({z_2})}}\,\ldots\,{e^{i\sqrt{m}\phi({z_N})}}
\,{e^{-i \,\int{d^2}z\,\sqrt{m}{\rho_0}\phi(z)}}
\rangle
\label{cblgs}
\ee
in which electrons are represented by
the operator ${e^{i\sqrt{m}\phi}}$. The last factor in the
correlation function  corresponds to a neutralizing
background ($\rho_0$ is the electron density); without it, 
this correlation function would vanish. Multi-quasihole
wavefunctions are obtained by inserting ${e^{i\phi/\sqrt{m}}}$
in this correlation function:
\bea
\label{cblqhs}
\langle\,{e^{{i\over\sqrt{m}}\phi({\eta_1})}}\ldots
{e^{{i\over\sqrt{m}}\phi({\eta_n})}}\,\,
{e^{i\sqrt{m}\phi({z_1})}}\,
{e^{i\sqrt{m}\phi({z_2})}}\,\ldots\,{e^{i\sqrt{m}\phi({z_N})}}
\,{e^{-i \,\int{d^2}z\,\sqrt{m}{\rho_0}\phi(z)}}
\rangle\br =\,\, {\prod_{\alpha>\beta}}{({\eta_\alpha}-{\eta_\beta})^{1/m}}
\,\,\,{\prod_{k,\alpha}}\,({z_k}-{\eta_\alpha})\,\,\,
{\prod_{i>j}}\, {({z_i} - {z_j})^m}~.
\eea

As may be seen directly from (\ref{laughgs}) or (\ref{cblgs}),
the electrons are fermions, as they must be. It may, furthermore, be seen
by inspection from (\ref{laughqhs}) or (\ref{cblqhs})
that a phase of ${e^{2\pi i}}=1$ is aquired when an electron
is taken around a quasiparticle. However, the advantage of the conformal
block construction is evident when we turn to the phase
aquired when one quasiparticle is taken around another.
According to (\ref{cblqhs}), this phase is $e^{2\pi i/m}$.
Whereas the $\eta$'s are merely parameters in an
electron wavefunction, so that their braiding properties must
be determined from the Berry's phase, the conformal blocks
put the $\eta$'s and $z$'s on an equal footing. {\it The key
conjecture is that the braiding properties of
both can be seen by inspection of the conformal
blocks}\cite{mooreread,blokwen,wilczek}.
These conformal blocks are isomorphic to the states of an abelian
Chern-Simons theory which describes these braiding
properties
\be
{\cal L} = m {a_\mu}{\epsilon^{\mu\nu\lambda}}
{\partial_\nu}{a_\lambda} + {a_\mu}{j^\mu}~
\ee
where $j^\mu$ is the quasihole current and
an electron is simply an aggregate of
$m$ quasiparticles.

In the $c=1$ theory, the electron is represented by
${e^{i\sqrt{m}\phi}}$; the quasihole, by ${e^{i\phi/\sqrt{m}}}$.
The primary fields of the algebra generated by the
Virasoro algebra together with the electron operator,
i.e. the rational torus, are
$1,{e^{i\phi/\sqrt{m}}},{e^{2i\phi/\sqrt{m}}},\ldots,
{e^{(m-1)i\phi/\sqrt{m}}}$.
These operators create excitations consisting of
$0,1,\ldots, m-1$ quasiholes. The primary
fields correspond to
the topologically inequivalent,
non-trivial excitations at $\nu=1/m$, 
since electrons
have trivial braiding properties with all excitations.
An excitation consisting of $k+m$ quasiholes is
equivalent to one
comprised of $k$ quasiholes because the additional
$m$ quasiholes have no effect on braiding, or,
in conformal field theory language, because
the former is
a descendent of the latter in the rational torus
algebra. Similarly, a quasiparticle is
equivalent to $m-1$ quasiholes. The $m$
different inequivalent conformal blocks
of the vacuum -- which correspond to the
degenerate quantum Hall ground states -- on the torus are constructed
via the Verlinde algebra by
creating a pair of conjugate fields, taking one around a loop,
and annihilating.

In the case of the Pfaffian state, a correspondence
can be made between the ground state and multi-quasihole
states and the conformal blocks of the $c = \frac{1}{2} +1$
conformal field theory. The braiding matrices, which
are embedded in a spinor representation of $SO(2n)$,
can be obtained from the latter. However, a direct check
cannot be done using the plasma analogy to compute the
Berry's phase matrix elements;
the more indirect arguments of \cite{wilczek}
must be used to justify the guess based on conformal field theory.
The $c=1$ part of the theory must be present in any quantum
Hall state; it simply `keeps track' of the charge. In
the wavefunction, it yields the Jastrow factors, which
determine the filling fraction. In the edge theory,
the $c=1$ sector of the theory describes the surface
density waves of an incompressible quantum Hall
droplet. The $c=\frac{1}{2}$ part of the theory is
responsible for the Pfaffian factor in the wavefunction,
and, hence, for the non-Abelian statistics. It also
describes the neutral fermionic excitations at
the edge of the Pfaffian state.

If we wish to follow the approach outlined in this section
to study the Haldane-Rezayi
state, we must, first, find a conformal field theory which
reproduces this state. As usual, there must be a
$c=1$ sector, which is the theory of a chiral boson,
with compactification radius $\sqrt{2}$
as would be expected as $\nu=\frac{1}{2}$.
According to \cite{mooreread,wenwu,milovan}, the pairing part
of the Haldane-Rezayi ground state is given
by a correlation function in the
$c = -2$ conformal field theory. We will discuss this at length
in the following section, but, for now, we state,
without justification, that there are dimension $1$
fermionic fields, $\partial\theta_\alpha$, in the
$c=-2$ theory and $\langle\partial{\theta_\alpha}
\partial{\theta_\beta}\rangle = -{\epsilon_{\alpha\beta}}/{z^2}$.
The electron can be represented as
${\Psi^{\rm el}} = \partial{\theta_1}{e^{i\sqrt{2}\phi}}u + 
\partial{\theta_2}{e^{i\sqrt{2}\phi}}v$ since
\bea
\langle{\Psi^{\rm el}}{\Psi^{\rm el}}\ldots{\Psi^{\rm el}}\rangle =
\langle\partial\theta\ldots\partial\theta\rangle\,\,
\langle{e^{i\sqrt{2}\phi}}\dots{e^{i\sqrt{2}\phi}}\rangle
= {\rm Pf}\left(\frac{{u_i}{v_j}-{v_i}{u_j}}{({z_i} - {z_j})^2}\right)
\,\,\,{\prod_{i>j}}{\left({z_i} - {z_j}\right)^2}
\eea

In the next section, we discuss the $c=-2$ conformal
field theory, with an eye towards calculating its
conformal blocks. In the following sections, we
use these conformal blocks to discuss the bulk
excitations of the Haldane-Rezayi state.

\section{Correlation Functions of the $c=-2$ Theory}

The $c=-2$ theory has been extensively studied (refs. \ct{Walgebra},
\ct{Gur}, \ct{Kauschcur}, \ct{Flohr},\cite{georgiev}).
Here we want to give a self-contained account
which includes all of the developments relevant to
our discussion of the Haldane-Rezayi state. Some of what we present 
here has not, to our knowledge, been published before.  

The $c=-2$ theory can be represented as a pair of 
ghost fields, or anticommuting
fields $\theta$, $\bar \theta$ with the action (ref. \cite{Gur})
\begin {equation}
S=\int \partial \theta \bar \partial \bar \theta
\end {equation}

This action has an $SU(2)$ (actually even an $SL(2,C)$) 
symmetry which becomes evident if we
introduce the `spin-up' and `spin-down' fields $\theta_1 \equiv \theta$
and $\theta_2 \equiv \bar \theta$ in terms of which the action is
\begin {equation}
\label{action}
S \propto \int \epsilon_{\alpha \beta} \partial 
\theta_\alpha \bar \partial \theta_\beta
\end {equation}  
where $\epsilon$ is the antisymmetric tensor.
Acting on $\theta$ by SU(2) matrices  does not change the action.
The SU(2) algebra is generated by the
SU(2) triplet of generators
\be
\label{walgebra}
W_{\alpha \beta} \propto \partial \theta_\alpha \partial^2 \theta_\beta +
                   \partial \theta_\beta  \partial^2 \theta_\alpha
\ee
of dimension 3, 
which form a ${\cal W}$-algebra rather than a Kac-Moody algebra (ref. 
\ct{Walgebra})\footnote{As has been noted in a number of publications,
the dimension 1 fields $\theta \partial \theta$ have logarithms in their
correlations functions and therefore do not form a Kac-Moody algebra.}.

The fields $\theta$ are complex. Nevertheless writing down the full action
\begin {equation}
\label{complex}
S \propto i \int  \epsilon_{\alpha \beta} \partial \theta_\alpha 
\bar \partial \theta_\beta - i \int
 \epsilon_{\alpha \beta} \partial \theta^\dagger_\alpha \bar \partial 
\theta^\dagger_\beta
\end {equation}
shows that $\theta^\dagger$ decouple from $\theta$ and we
can consider them independently. If, on the other hand, we include
them, the central charge for the theory
\rf {complex} is $c=-4$. We emphasize that $\bar \theta$ is
not a complex conjugate of $\theta$, but is just another field.  
Alternatively, we could take $\theta$, $\bar \theta$ to
be real fields with an $SL(2,R)$ symmetry.

To quantize the theory \rf{action} we have to compute the fermionic
functional integral
\be
\label{Ferm}
\int {\cal D} \theta {\cal D} \bar \theta \exp (-S)
\ee
We note that computed formally this fermionic path integral
is equal to zero due to the ``zero modes'' or constant parts of
the fields $\theta$ which do not enter the action \rf{action}. 
To make it nonzero we have to insert the fields $\theta$ into the
correlation functions (compare with ref. \ct{Friedan}), as in
\be
\int {\cal D} \theta {\cal D} \bar \theta \, \, \bar \theta (z) 
\theta(z) \exp (-S)
= 1
\ee
Therefore, the vacuum $|0\rangle$
of this theory is somewhat unusual. Its norm
is equal to zero,
\be
\left\langle 0 | 0 \right \rangle = 0
\label{strange1a}
\ee 
while the explicit insertion of the fields $\theta$ produces nonzero
results
\be
\label{strange1b}
\VEV{\bar \theta(z) \theta(w)}=1
\ee

Furthermore, if we want to compute correlation functions of the
fields $\partial \theta$ we also need to insert the zero modes
explicitly,
\bea
\label{strange2}
\VEV{\partial \theta(z) \partial \bar \theta(w) } = 0, \ \ {\rm but} \br
\VEV{\partial \theta(z) \partial \bar \theta(w) \bar \theta(0)  \theta(0)} =
- {1 \over (z-w)^2}
\eea
The second correlation function is computed by analogy to the free
bosonic field. 

  From the point of view of conformal field theory, the 
strange behavior of \rf{strange1a}, \rf{strange1b}, and \rf{strange2}
can be explained in terms of the logarithmic operators which
naturally appear at $c=-2$. As was discussed in \ct{Gur}, the
theory $c=-2$ must necessarily possess an operator $\t{I}$ of dimension
$0$, in addition to the unit operator $I$, such that
\be
\label{jacobi}
[L_0, \t{I}] = I
\ee
(where $L_0$ is the Hamiltonian).
Moreover, it can be proved by general arguments of conformal
field theory, such as conformal invariance and 
the operator product expansion, that 
the property \rf{jacobi} necessarily leads to the correlation functions
(refs. \ct{Gur} and \ct{Kogan})
\bea
\label{Kogan}
\VEV{I I}=0, \br
\VEV{I(z) \t{I}(w)}= 1, \br
\VEV{\t{I}(z) \t{I}(w)} = - 2 \log(z-w)
\eea
These relations force us to conclude that the operator $\t{I}$
must be identified with the normal ordered product of $\theta$ and
$\bar \theta$ \ct{Zamol},
\be
\t{I} \equiv  - :\theta \bar \theta: = - {1 \over 2}
\epsilon_{\alpha \beta} \theta_\alpha \theta_\beta
\ee

The stress energy tensor of the theory \rf{action} is given by
\be
T=:\partial \theta \partial \bar \theta:
\ee
and it is easy to see that its expansion with $\t{I}$ is indeed given
by
\be
\label{expansion}
T(z) \t{I}(w)= {1 \over (z-w)^2} + {\partial \t{I} \over z-w} + \dots
\ee

The mode expansion of the fields $\theta$ has to be written in the
form
\be
\label{modes}
\theta(z) = \sum_{n \not = 0} \theta_n z^{-n} + \theta_0 \log(z) + \xi
\ee
where $\xi$ are the crucial zero modes (they disappear in the
expansion for $\partial \theta$). Here $n \in  Z$ in the
untwisted sector (ie. with periodic boundary conditions)
and $n \in Z + \frac{1}{2}$ in the twisted sector
(anti-periodic boundary conditions).

To be consistent with the earlier results \rf{strange2} and \rf{expansion}
we have to impose the following anticommutation relations (compare with ref.
\ct{Kauschcur})
\bea
\label{anticom}
\left\{ \theta_n, \bar \theta_m \right\} = 
{1 \over n} \delta_{n+m,0}, \ n \not = 0 \br
\left\{ \theta_0, \bar \theta_0 \right\} = 0 \br
\left\{ \theta_m, \theta_n \right\} = \left\{ \bar \theta_n, \bar
\theta_m \right\} = 0 \br
\left\{ \xi, \bar \xi \right\}= 0 \br
\left\{ \xi, \bar \theta_0 \right\}= 1 \br
\left\{ \theta_0, \bar \xi \right\} = -1
\eea

The last two relations
are absolutely crucial in keeping 
\rf{expansion} intact. The mode expansion $\theta_n$ should
not be confused with the notations $\theta_1$ and $\theta_2$
introduced earlier. To avoid confusion we will primarily use
the $\theta, \bar \theta$ notation.

Note that the modes $\xi$ become the creation operators for logarithmic
states. Indeed, 
\be
\theta_n | 0\rangle =0\,\,{\rm for} \,\, n \ge 0
\ee
and
\be
\t{I} |0 \rangle = \bar \xi \xi | \rangle
\ee  

The mode expansion \rf{modes} together with \rf{anticom} 
and 
\be
\VEV{0|0}=0, \ \VEV{\bar \xi \xi }=1
\ee
can be
used to compute any correlation function in the theory.

For instance, we can reproduce the correlation functions of
\rf{Kogan}
\be
\VEV {I(z) \t{I}(w) } = \VEV {\bar \theta(w) \theta(w)} =
\VEV{\bar \xi \xi} = 1 
\ee
while
\bea
\label{Kogan1}
\VEV{\t{I} (z) \t{I} (w)} = \VEV{ :\bar \theta(z) 
\theta(z): :\bar\theta(w)  
\theta(w): } = \br
\VEV{ \bar \xi \theta(z) \bar \theta (w) \xi } + \VEV { \bar \theta(z) \xi 
\bar \xi \theta(w) } = - 2 \log(z-w)
\eea
The last line of \rf{Kogan1} can be computed either directly 
in terms of modes
or by comparison with \rf{strange2}.

As has been discussed at length in the literature, the fields $W$ introduced
in \rf{walgebra} form a ${\cal W}$-algebra and in fact all the states
of the $c=-2$ theory can be classified according 
to various representations of that algebra. A clear review can be found in
ref. \ct{Kausch}. Six representations are listed in that paper. They 
can easily be represented in terms of the fields of our theory. 
We have the unit operator $I$, the logarithmic operator $\t{I} = - : \theta 
\bar \theta:$, the SU(2) doublet of dimension 1 fields $\partial \theta$ and
$\partial \bar \theta$, the twist field $\mu$ of dimension $-1/8$,
a doublet of twist fields 
$\sigma_{\alpha}\equiv {\left( \theta_{\alpha} \right)}_{- {1 \over 2}} \mu$ 
of dimension\footnote{$\theta_{-{1\over 2}}$
is the mode expansion \rf{modes} for $\theta$ where $n\in Z+{1 \over 2}$
to reproduce the twisted sector. The zero modes are naturally absent
in that sector.} $3/8$, 
and finally a structure of fields $\theta$, $\partial \theta$ and
$\theta \partial \theta$ connected with each other by the action
of the Virasoro generators $L_n$.

With all the preliminaries completed we can proceed to construct 
the correlation functions of the fields $\theta$. The correlation
function
\be
\label{Haldane}
\VEV {\partial \theta(z_1) \partial \bar \theta(w_1) \dots
\partial \theta(z_n) \partial \bar \theta(w_n) \t{I} } =
\sum_{\sigma} {\rm sign}{\sigma} \prod_{i=1}^n {1 \over (z_i - 
w_{\sigma(i)})^2},
\ee
where $\sigma(i)$ is the permutation of the numbers $1$, $2$, $\dots$, $n$,
reproduces the Haldane-Rezayi wave function. 
Note the explicit insertion of the logarithmic operator $\t{I}=:\bar
\theta \theta:$ to make \rf{Haldane} nonzero. For convenience,
we express the correlation functions in this section
in `$z$-$w$' notation in which the $\theta$'s are at the
points $z_i$ and the $\bar \theta$'s are at the $w_i$'s.
In the next section, we revert to the `$u$-$v$' notation
which is better adapted for a discussion of wavefunctions.

The correlation functions in the twisted sector can be found 
by splitting the logarithmic operator into two twist fields $\mu$
according to the general formula (ref. \ct{Gur})
\be
\mu(z) \mu(w) \approx I \log (z-w) + \t{I}
\ee
and is equal to 
\bea
\label{twisted}
\VEV {
\partial \theta(z_1) \partial \bar \theta(w_1) \dots
\partial \theta(z_n) \partial \bar \theta(w_n)
\mu(\eta_1) \mu(\eta_2) }  = \br
{\left( \eta_1-\eta_2 \right)}^{1 \over 4} \sum_{\sigma} 
{\rm sign\sigma}\prod_{i=1}^n { 
(z_i-\eta_1)(w_{\sigma(i)}-\eta_2)+(z_i-\eta_2)(w_{\sigma(i)}-\eta_1) \over
(z_i-w_{\sigma(i)})^2 \sqrt{ (z_i-\eta_1) (z_i-\eta_2)(w_{\sigma(i)}-\eta_1)
(w_{\sigma(i)}-\eta_2)} }
\eea
Note that we do not need the logarithmic operator any more. It has
been split into two twist fields. Alternatively, we can say that in
the twisted sector the summation in \rf{modes} is over half integer numbers
 and the zero modes no longer
enter the expansion for the fields $\theta$. 

Correlation functions of the type
\rf{twisted} are, as we will see below, useful for
constructing the bulk excitations. 
However, the twist fields are not the only way of 
doing it. We could also split the logarithmic
operator according to the
operator product expansion
\be
\t{I}(z) \t{I}(w) = -2 \log(z-w) \t{I} + \dots
\ee
which follows from \rf{Kogan}. 
The following correlation function
\be
\VEV{
\partial \theta(z_1) \partial \bar \theta(w_1) \dots
\partial \theta(z_n) \partial \bar \theta(w_n)
\t{I} (u_1) \t{I}(u_2) }
\ee
will be needed in the next section. It can be computed by either solving the
differential equations of conformal field theory, or by the straightforward
mode expansion \rf{modes} and \rf{anticom}. Either method results in
\bea
\label{newpart}
\VEV{
\partial \theta(z_1) \partial \bar \theta(w_1) \dots
\partial \theta(z_n) \partial \bar \theta(w_n)
\t{I} (u_1) \t{I}(u_2) } = \br - 2 \log(u_1-u_2) 
\sum_{\sigma} {\rm sign}{\sigma} \prod_{i=1}^n {1 \over (z_i - 
w_{\sigma(i)})^2} - \br
\sum_{\sigma} {\rm sign}{\sigma} \sum_{k=1}^{n} \left\{
\prod_{i \not = k} \left( {1 \over (z_i -
w_{\sigma(i)})^2 }\right)  {(u_1-u_2)^2 \over (u_1-z_{k}) (u_1-w_{\sigma(k)})
(u_2-z_k) (u_2-w_{\sigma(k)})} \right\}
\eea

We see that it splits into two terms. One is the product of the
Haldane-Rezayi wave function \rf{Haldane} and the logarithm. The other
is a nontrivial expression. In fact, it is easy to get rid of the
trivial part by taking one of the logarithmic operators to infinity.
In doing so we have to remember the transformation law for the
logarithmic fields which follows from \rf{jacobi},
\be
\t{I}(f(z))=\t{I} (z) + \log \left( {\partial f \over \partial z } \right)
\ee
According to the standard procedure, taking the position
of the field $\t{I}(z)$ to infinity 
corresponds to taking the position of the
field $\t{I}(1/z)=\t{I}(z) - 2 \log(z)$ to the origin.
Therefore the trivial part of \rf{newpart} 
disappears.

\section{Topological Properties of
Bulk Excitations in the Haldane-Rezayi State.}

Armed with the preceding results, we can discuss the bulk
excitations of the Haldane-Rezayi state. The discussion
will be more complicated than but otherwise directly analogous
to the discussion of the Laughlin state in section III and
of the Pfaffian state in \cite{wilczek}.

The primary fields of the $c=-2+1$ theory are:
$1,{e^{i\phi/\sqrt{2}}}$, which create the ground state and
the Laughlin quasiparticle; $\partial\theta,
\partial\theta{e^{i\phi/\sqrt{2}}}$, which create
neutral fermions in the bulk;
$\mu{e^{i\phi/2\sqrt{2}}},
\mu{e^{i\phi/2\sqrt{2}+i\phi/\sqrt{2}}}$,
${\sigma_\alpha}{e^{i\phi/2\sqrt{2}}},
{\sigma_\alpha}{e^{i\phi/2\sqrt{2}+i\phi/\sqrt{2}}}$,
which create half flux quantum quasiholes;
and
$\t{I}{e^{i\phi/\sqrt{2}}}, \t{I}{e^{i\phi\sqrt{2}}}$.
Although there are $10$ fields, they are not
obtained by simply multiplying the $2$ primary fields
of the $c=1$ theory with the $5$ of the $c=-2$
theory. The last $3$ pairs of fields involve particular
combinations of fields from the $c=-2$ and $c=1$
theories. These are necessary to give wavefunctions
which are single-valued in the electron coordinates.
Milovanovic and Read have shown that this requirement is
equivalent to an orbifold construction \cite{milovan}. These $10$
primary fields correspond to the $10$ topologically distinct
bulk excitations of the Haldane-Rezayi state. The corresponding
wavefunctions are ($p,{p_\alpha}=0,1$):
\be
{\Psi_I}\,\, =\,\,
{\rm Pf}\left(\frac{{u_i}{v_j}-{v_i}{u_j}}{({z_i} - {z_j})^2}\right)
\,\,\,{\prod_i}{\left({z_i}-\eta\right)^p}
\,\,\,{\prod_{i>j}}{\left({z_i} - {z_j}\right)^2}\,\,\,
\label{hrsi}
\ee
\be
{\Psi_{\partial\theta}}\,\, =\,\,
{\cal A}\left(\frac{\chi_1}{(\eta-{z_1})^2}\,
\frac{{u_2}{v_3}-{v_2}{u_3}}{({z_2} - {z_3})^2}\,\,\ldots\,\right)
\,\,\,{\prod_i}{\left({z_i}-\eta\right)^p}
\,\,\,{\prod_{i>j}}{\left({z_i} - {z_j}\right)^2}\,\,\,
\label{hrstheta}
\ee
\be
{\Psi_\mu}\,\, =\,\,
{\left({\eta_1}-{\eta_2}\right)^{3/8}}
{\rm Pf}\left(\frac{\left({u_i}{v_j}-{v_i}{u_j}\right)\,\,
\left(\left({z_i} - {\eta_1}\right)
\left({z_j} - {\eta_2}\right) + i\leftrightarrow j\right)}
{({z_i} - {z_j})^2}\right)
{\prod_{i,\alpha}}{\left({z_i}-{\eta_\alpha}\right)^{p_\alpha}}
\,\,\,\,{\prod_{i>j}}{\left({z_i} - {z_j}\right)^2}
\label{hrsmu}
\ee
\bea
\label{hrssigma}
{\Psi_\sigma}\,\, =\,\,
{\left({\eta_1}-{\eta_2}\right)^{19/8}}\,
{\cal A}\Biggl(
\frac{\left({u_1}{v_2}+{v_1}{u_2}\right)\left({z_1}-{z_2}\right)}
{\left({\eta_1}-{z_1}\right)\left({\eta_2}-{z_1}\right)
 \left({\eta_1}-{z_2}\right)\left({\eta_2}-{z_2}\right)}
\,\times\br
\,\,\frac{\left({u_3}{v_4}-{v_3}{u_4}\right)\,\,
\left(\left({z_3} - {\eta_1}\right)
\left({z_4} - {\eta_2}\right) + 3\leftrightarrow 4\right)}
{({z_3} - {z_4})^2}\,\,\ldots\,\Biggr)
{\prod_{i,\alpha}}{\left({z_i}-{\eta_\alpha}\right)^{p_\alpha}}
\,\,\,\,{\prod_{i>j}}{\left({z_i} - {z_j}\right)^2}
\eea
\bea
\label{hrslog}
{\Psi_{\t{I}}}\,\, =\,\,
{\cal A}\left(\frac{\left({u_1}{v_2}-{v_1}{u_2}\right)
{\left({\eta_1}-{\eta_2}\right)^2}}
{\left({z_1}-{\eta_1}\right) \left({z_1}-{\eta_2}\right)
\left({z_2}-{\eta_1}\right) \left({z_2}-{\eta_2}\right)}
\,\,\frac{{u_3}{v_4}-{v_3}{u_4}}
{({z_3} - {z_4})^2}\,\,\ldots\,\right)\times\br
{\prod_{i,\alpha}}{\left({z_i}-{\eta_\alpha}\right)^{{p_\alpha}+1}}
\,\,\,\,{\prod_{i>j}}{\left({z_i} - {z_j}\right)^2}
\eea
The states (\ref{hrstheta}), (\ref{hrssigma})
are not legitimate lowest Landau level wavefunctions.
However, they have the correct braiding properties, and
should be thought of as shorthand for the correct wavefunctions
which can be constructed along the lines of the neutral
fermion wavefunctions given in \cite{greiter}. If this is done
for the state (\ref{hrssigma}), it will vanish,
but when there are more quasiholes,
there are non-trivial conformal blocks with $\sigma$'s
which are different from those with $\mu$'s (see below). In
(\ref{hrsmu})-(\ref{hrslog}),
we have created pairs of excitations, as must be done on
the sphere. In the plane, single excitation wavefunctions
can be obtained by taking
${\eta_2}\rightarrow\infty$.\footnote{As a general rule,
conformal field theory imposes requirements such as
charge neutrality, flux quantization, etc. which
must be satisfied by wavefunctions on the sphere. These
can be relaxed on the plane by taking some of the
quasiparticles to infinity. The conformal blocks must
also be spin-singlets, which means that they must
be invariant under rotations of the $\sigma_\alpha$'s
and the $\partial\theta_\alpha$'s. Taken as electron
wavefunctions, however, they need not be singlets
under rotations of the electron spins alone.}
The excitations given by
(\ref{hrsi})-(\ref{hrssigma}) have analogs in
other paired states. However, (\ref{hrslog}),
which raises the ground state degeneracy on the torus
to $10$, rather than $6$ or $8$ as it is in the
Pfaffian and $(3,3,1)$ states, is new. There is an
analogous wavefunction in the $(3,3,1)$ state which,
as in the Haldane-Rezayi state,
can result from bringing together (i.e. fusing)
two half-flux quantum quasiholes. However, it has
trivial braiding properties, and therefore does
not contribute to the ground state degeneracy of the $(3,3,1)$
state on the torus. In the Haldane-Rezayi state,
however, the $\t{I}$ excitation (\ref{hrslog}) braids
non-trivially with the half-flux quantum quasiholes.

Non-Abelian statistics first raises its
head when there are four quasiholes.
Unfortunately, we
cannot explicitly calculate the corresponding conformal blocks,
which would require calculating
$\langle \mu\mu\mu\mu\partial\theta\ldots\partial\theta\rangle$
and similar conformal blocks with more twist fields.
Since $\mu \mu \sim I + \t{I}$, there are $2^{n-1}$ conformal
blocks with $2n$ $\mu$'s (and any number of $\partial\theta$'s).
To count the other $2n$ quasihole states,
we have to count all conformal blocks with $2n$ fields which can be either
$\mu$ or $\sigma_{\alpha}$. Recall that $\sigma_\alpha$ is obtained by 
fusing $\mu$ and $\partial\theta_\alpha$. Hence, if we call a half flux
quantum quasihole operator $s^a$, which could be either $s^0 = \mu$ or
$s^{\pm\frac{1}{2}} = \sigma_{\alpha}$, and if we further write
$I^a, \tilde{I}^a$ to denote members of the conformal families of
$I,\tilde{I}$ for $a\in Z$, and of $\partial\theta_{\alpha}, 
\partial\theta_{\alpha}\tilde{I}$ (the spin
doublet of conformal weight $h=1$) for $a\in Z + 
\frac{1}{2}$, we can collect all fusion rules in the compact form
$[s^a] \times [s^b] = [I^{a+b}] + [\tilde{I}^{a+b}]$ and
$[s^a] \times [I^b] = [s^a] \times [\tilde{I}^b] = [s^{a+b}]$.
Conformal blocks as 
$\langle s^{a_1}(z_1)s^{a_2}(z_2)\ldots s^{a_{2n}}(z_{2n})\rangle$ have
an essentially predetermined form with some
straighforward products of powers of $(z_i - z_j)$ and products of
certain functions depending an all possible crossing ratios.
However, these functions depend only on the conformal weights of the
fields in the correlator and the internal channels of the conformal block, 
not on the spin indices directly.

Let us assume momentarily that we work in a basis where the metric on the
internal channels is diagonal, so that we don't have to think about
additional indices for the endpoints of internal propagaters.
What this means is that we only have to keep track of the spin
indices modulo integers. Then, counting conformal blocks is very
simple. With $2n$ fields $s^{a_i}(z_i)$ in a correlator we have $(n-1)$
internal chanels of type $I^a$ or $\tilde{I}^a$, the other internal
chanels being of type $s^{a}$. Since each of the former internal chanels
can be either $I^a$ or $\tilde{I}^a$, we have $2^{n-1}$ possible choices.
Further, each such internal channel has $a \equiv 0$ modulo integers, or
$a \equiv \frac{1}{2}$ modulo integers (the outer chanels, i.e.\ fields,
appropiately chosen). Due to the overall condition that in total we need 
spin $0$ fixes the $(n-1)$-th internal chanel, if the other (n-2) are chosen.
So we get in total
$2^{n-1}2^{n-2} = 2^{2n-3}$ possible conformal blocks.

We turn now to the monodromy matrices which are generated when
quasiholes are taken around one another. Consider the simplest
non-trivial case, with four quasiholes. The two degenerate
states can be obtained from the conformal blocks of
$\langle \mu\mu\mu\mu\partial\theta\ldots\partial\theta\rangle$
(correlation functions with some of the $\mu$'s replaced
by $\sigma$'s have identical conformal blocks in
the four-quasihole case, which is the simplest
instance of the above reduced degeneracy). Even in the
absence of the explicit forms of these conformal blocks,
the monodromy matrices can be obtained from the differential equations
which the conformal blocks satisfy (they are the equations for the
full elliptic integrals, see \ct{Gur}).
Apart from a trivial phase $e^{-\pi i/4}$ which
arises from the chiral boson sector of the theory, the
monodromies are:
\be
\left(\matrix{1& 0\cr -2 i &  1}\right)
\label{mono0}
\ee
when $\eta_1$ is taken around $\eta_2$,
\be
\left(\matrix{1& - 2 i\cr   0& 1}\right)
\label{mono1}
\ee
when $\eta_1$ is taken around $\eta_4$, and
\be
\left(\matrix{-3 & 2 i\cr 2 i&  1}\right)
\label{monoinf}
\ee
when $\eta_1$ is taken around $\eta_3$. ${\eta_1},{\eta_2},{\eta_3},
{\eta_4}$ are the positions of the four quasiholes and
the matrices refer to the preferred basis of four-quasihole
states which is given by the conformal blocks (which, again,
we are unable to obtain explicitly). The most salient
property of these matrices is that they are not unitary.
However, they are unitary with respect to the indefinite metric
\be
\left(\matrix{0& 1\cr 1& 0}\right)
\ee
It is worth noting that this metric is precisely 
the metric on the space of 4-point conformal blocks
which is necessary to obtain single-valued correlation
functions from the left and right conformal blocks
(see \cite{Gur}).
It is natural to conjecture that these $SU(1,1)$ monodromy matrices
are the analytic continuation to this indefinite
metric of the correct $SU(2)$ monodromy matrices of the Haldane-Rezayi
quasiholes, which must be unitary with respect to the
definite metric of the Hilbert space of lowest Landau level
states. This analytic continuation may be done for
(\ref{mono0})-(\ref{monoinf})
but it is not enlightening. In \cite{wilczek},
it was found that the monodromy
matrices of the Pfaffian state are given by
certain $SO(2n)$ rotations in their spinor
representation. It is plausible that the
non-abelian statistics of the $2n$ quasihole states
of the Haldane-Rezayi state is given by a
reducible representation of some group $G$ (and,
hence, of the braid group) because states of different
spins will not mix. One possibility is that the
states can be grouped into a direct sum of the $SO(2n)$
or $SU(n)$ irreducible representations into which a product of
$SO(2n)$ spinor representations may be decomposed.

\section {Relationship Between $c=-2$ and $c=1$ theory.}

A number of papers have
established a relationship between the 
the partition functions of 
$c=-2$ and $c=1$, $R=1$ theories (ie. the $c=1$ theory of a Dirac
fermion or a free boson with compactification radius
$R=1$)\ct{Flohr}, \ct{Kausch}.  
While the proof requires a careful construction of $c=-2$
characters taking into account the presence of the logarithmic
operators (ref. \ct{Flohr}), there is a way to roughly understand
the relationship in rather simple terms (see, especially, \cite{ludwig}). 

One can easily check that for each operator of dimension $h_{c=-2}$ 
in the $c=-2$ theory there is an operator of dimension $h_{c=1}$ in the
$c=1$, $R=1$ theory such that
\be
h_{c=-2} - {-2 \over 24} = h_{c=1}- {1 \over 24}
\ee
Therefore, in the theory on the cylinder, where
the partition function is computed, and where the edge theory lives, 
their zero-point energies are the same. 

Moreover, for each descendant state in the $c=-2$ theory there is a 
corresponding state
in the $c=1$, $R=1$ theory. 
Indeed, take the latter theory as represented
by a Dirac fermion
\be
S= \int \psi^\dagger \bar \partial \psi
\ee
The modes $\psi^\dagger_{n}$ and $\psi_{-n}$, $n>0$ can be used
to construct descendant states,
\be
{\psi^\dagger_{n_1}}\ldots{\psi^\dagger_{n_\alpha}}\,\,
{\psi_{-m_1}}\ldots{\psi_{-m_\beta}}\,|0\rangle
\label{dfs}
\ee
which have the same
energies and momenta as the states
created by $\theta_{-n}$ and $\bar \theta_{-n}$,
\be
{\theta_{-n_1}}\ldots{\theta_{-n_\alpha}}\,\,
{{\bar\theta}_{-m_1}}\ldots{{\bar\theta}_{-m_\beta}}\,|0\rangle
\ee

The mode expansion of the field $\psi$ (on the plane) is
\be
\psi(z) = \sum_{n} \psi_{n} z^{-n-{1 \over 2}}
\ee
so the Dirac fermion has half-integral momenta in the untwisted
sector and integral momenta in the twisted sector,
while the opposite is true for the $c=-2$ theory.
Therefore, we map the twisted sector
of $c=-2$ into the untwisted one of $c=1$ and {\sl vice versa}.
(See \cite{ludwig} for a more detailed discussion of the
mapping at the level of the mode expansions.)

Of course, there is still a question of the zero modes $\xi$;
they do not seem to correspond to anything in $c=1$. However,
the zero modes $\xi$ are rather special fields. They must
be present as out- (or in-) states of the $c=-2$ theory to make the
correlators of the theory nonzero, and
they have to be present only once (since $\xi^2=0$).

As far as many of their conformal properties are
concerned, the theories of anticommuting bosons and of Dirac fermions
are the same on the cylinder. 
Their respective vacua have the same energy and for each operator
of $c=-2$ theory there exists an operator at $c=1$, $R=1$. 
However, the higher modes of the energy-momentum tensor
are different in the two theories. Consequently, the Virasoro
algebra representations are different; in the $c=1$ theory, they
are unitary while in the $c=-2$ theory they are non-unitary.

\def\Vm{{\cal V}_{-1/8}}
\def\Vp{{\cal V}_{3/8}}
\def\Vn{{\cal V}_{0}}
\def\Ve{{\cal V}_{1}}
\def\Wn{{\cal W}_{0}}
\def\We{{\cal W}_{1}}
\def\Rn{{\cal R}_{0}}
\def\Re{{\cal R}_{1}}
To understand the equivalence of the $c=-2$ and $c=1$ theories better, 
let us take a closer look at the sectors of the $c=-2$ theory.
Ordinarily, each primary field $\phi(z)$ of a conformal field theory 
and all its descendants generate a highest weight representation
of the Virasoro algebra, 
perhaps with a chiral symmetry algebra (see rf. \ct{BPZ}).
Moreover, that representation is irreducible. We achieve its irreducibility
by removing all the descendants of the primary field which are themselves
primary operators.
The Hilbert space of a conformal
field theory can then  be written as a direct sum over irreducible
highest weight representations.

The problem we immediately encounter in the $c=-2$ theory
is that its states do not constitute ordinary irreducible highest
weight representations of the maximally extended chiral symmetry algebra
(${\cal W}$-algebra)
or even of the Virasoro algebra itself.
We know that we have a state there, $|\t{I}\rangle$, such that $L_0 |\t{I}
\rangle = |0\rangle$, 
$|0\rangle$ being the vaccuum while $|\t{I}\rangle = \t{I}(0) |0 \rangle$. 
$|\t{I}\rangle$ and $|0\rangle$ 
have to be considered together,
and together they are usually 
referred to as  a reducible but indecomposable representation (ref. 
\ct{KauschGab}).
Indeed, we can certainly reduce it by considering a subset of it, consisting
of $|0\rangle$ and its descendants only, without $|\t{I}\rangle$. 
However, we cannot
decompose it into a direct product of $|0\rangle$ and $|\t{I}\rangle$ 
as $L_0 |\t{I}\rangle = 
|0\rangle$.  

According to \ct{Kausch}, \ct{KauschGab}, \ct{Rohsiepe}
there are four sectors generated by operators with $h_{c=-2}\in\{-\frac{1}{8},
0,\frac{3}{8},1\}$, denoted $\Vm,\Rn,\Vp,\Re$ in the following (we use the
notation of \ct{Kausch}) and the characters of 
these representations are
\bea
\label{characters}
\chi_{\Vm} = \frac{1}{\eta(\tau)}\Theta_{0,2}(\tau) \br
\chi_{\Vp} = \frac{1}{\eta(\tau)}\Theta_{2,2}(\tau) \br
\chi_{\Rn} = \chi_{\Re} = \frac{2}{\eta(\tau)}\Theta_{1,2}(\tau)
\eea
where $\eta(\tau) = q^{1/24}\prod_{n>0}(1-q^n)$ is the Dedekind
eta function, $\Theta_{\lambda,k} = \sum_{n\in Z}q^{(2kn+\lambda)^2/4k}$
are ordinary Theta functions, and $q=\exp(2\pi i\tau)$ is the modular
parameter of the torus.

Note the multiplicity of 2 in the last two characters. It forces an
overall multiplicity of 4 in the diagonal partition function 
to ensure modular invariance\footnote{Moular invariance of the torus
partition function of a conformal field theory is an important requirement
for consistency. In the context of the theory of the bulk of
a quantum Hall state, it is just the statement that the theory,
when put on the torus, should be independent
of the coordinate system on the torus.
It has been proven \ct{Nah91} that conformal invariance of a 
theory on $S^2$ implies modular invariance on the torus, if $L_0$ is diagonal. 
This should extend to the case of logarithmic conformal field theory by the
limiting procedure described in \ct{Flohr}.}
%\bibitem[Nah91]{Nah91} {\sc W. Nahm},
%  {\em A proof of modular invariance},
%   Int. J. Mod. Phys. {\bf A6} (1991) 2837, in Proc. Trieste July 1990 {\em
%   Topological methods in quantum field theories}, World Scientific, 1991
\be
\label{partition1}
Z_{c=-2} = |\chi_{\Rn}|^2 + |\chi_{\Re}|^2 + |2\chi_{\Vm}|^2 + |2\chi_{\Vp}|^2
         = 4Z_{c=1}(R=1)
\ee
such that equivalence of the partition functions of the $c=-2$ theory and
the $c=1$ theory is really established only up to a factor of 4. Moreover,
there is no way to avoid the multiplicities of the $\Vm$ and $\Vp$ 
representations. The overall multiplicity of 4 stems from the zero modes
$\xi,\bar{\xi}$. It turns out that both indecomposable representations
are formed out of four subsectors according to the four possible ways to
distribute these zero modes. However, the combinatorics of the subsectors
falls into just two different types which coincide with the combinatorics
of the irreducible subrepresentations of $\Rn$ and $\Re$, called $\Vn$ and
$\Ve$ respectively. Their characters are \ct{Flohr}, \ct{Kausch}
\bea
\label{newchars}
\chi_{\Vn}=\frac{1}{2\eta(\tau)}\left(\Theta_{1,2}(\tau)+\eta^3(\tau)\right)\br
\chi_{\Ve}=\frac{1}{2\eta(\tau)}\left(\Theta_{1,2}(\tau)-\eta^3(\tau)\right)\br
\eea
and each of these two sector types appears twice in each of the indecomposable
representations. We thus conclude that the partition functions consists
of four copies of the $c=1$ Dirac fermion partition function, one for each
possible combination of the $\xi,\bar{\xi}$ zero modes.

Although we don't need to take this multiplicity into account on the $c=1$
side, because there everything factorizes, this multiplicity is intrinsic
on the $c=-2$ side due to the fact that some representations are 
indecomposable. However, there are some disadvantages with this approach to
the $c=-2$ theory: The modular behavior of the characters \rf{characters} is
ambigous due to the equivalence of $\chi_{\Rn}$ and $\chi_{\Re}$. Moreover,
the $S$-matrix for the modular transformation $S:\tau\mapsto -1/\tau$ does not
reproduce the correct fusion rules via the Verlinde formula.

In \ct{Flohr} it was attempted to overcome these difficulties by using the
fact that the $\xi,\bar{\xi}$ zero modes are necessary to make any $n$-point
function non-zero. That means that there is a way to partially factorize
the untwisted part of the partition function by splitting each indecomposable
representation into its irreducible subrepresentation and the part with
the opposite $\theta$-fermion number (the total fermion number including the
$\xi$ zero mode is always even in $\Rn$ and odd in $\Re$). The result is
\be
\tilde{Z}_{c=-2} = |\chi^{}_{\Vm}|^2 + |\chi^{}_{\Vp}|^2 + \left(
                   \chi^{}_{\Vn}\chi^*_{\Wn} + 
                   \chi^{}_{\Ve}\chi^*_{\We} + c.c.
                   \right) = Z_{c=1}(R=1)
\ee
where $\chi_{\Wn} = \chi_{\We} = \Theta_{1,2}(\tau)/\eta(\tau)$.
The non-diagonal structure precisely resembles the non-diagonal structure
of the conformal blocks necessary in the $c=-2$ theory to ensure crossing
symmetry and single valuedness of the four point function, see \ct{Gur}.

We conclude by mentioning that this partition function is certainly modular
invarinat, but the set of characters $\{\chi_{\Vm},\chi_{\Vp},\chi_{\Vn},
\chi_{\Wn},\chi_{\Ve},\chi_{\We}\}$ is not. One of the results of
\ct{Flohr} is that by introducing a regularizing term 
$\pm i\alpha\log(q)\eta^3(\tau)$ into $\chi_{\Wn},\chi_{\We}$, one recovers
modular covariance for the characters. However, the physical meaning of
a $\log(q)$ term in character functions remains unclear. As long as $\alpha$
is taken non-zero, one has a well-defined $S$-matrix which can be used to
calculate fusion coefficients via the Verlinde formula. As shown in
\ct{Flohr}, the latter have physical meaning only in the limit $\alpha
\rightarrow 0$ and coincide then with explicit results.

\section{Edge Theory of the Haldane-Rezayi State}

The preceeding considerations inspire us to hope
for the following happy ending to our story:
the neutral sector of the low-energy
edge theory of the Haldane-Rezayi
state is a $c=1$ Dirac fermion.

How can we show that this assertion is correct?
Since a quantum-mechanical theory is defined by
its Hilbert space of states, inner product,
and algebra of observable operators, we must show that these
structures are identical for the $c=1$ theory and
the edge excitations annihilated by the
Hamiltonian (\ref{halrezham}). Clearly, the Hilbert
spaces, (\ref{hres}) and (\ref{dfs}),
are the same.\footnote{Almost. The twisted sector of the
$c=1$ Dirac theory has $2$ zero modes, while the edge excitations
of the Haldane-Rezayi state begin at angular momentum $1$,
ie. $k=2\pi/L$. This zero mode must be projected out of the theory,
which can be done very naturally in the truncation from
a $c=2$ theory described below.}
The spectra (assuming
that the energy is proportional to the angular momentum, as before)
and, hence, the partition functions are, as well. Of course,
the same may be said for the $c=-2$ theory (ignoring
subtleties associated with the zero modes, $\xi$, $\bar \xi$),
as we discussed in the previous section. The observables --
such as the local energy and spin densities -- and the inner
product must distinguish the correct edge theory. However,
these are difficult to calculate.

In trying to calculate the inner products of the edge
excitations (\ref{eswf}), we run into a familiar roadblock: in the absence
of a plasma analogy, there is no painless way of
doing this calculation. This complicates matters when we
turn to the algebra of observables, because we are interested
in these operators {\it projected into the low-energy
subspace}. If this were simply a matter of projecting into the
lowest Landau level, it would be no problem. However,
we must project into the zero-energy subspace of the
Hamiltonian (\ref{halrezham}), since this is the subspace which
contains the low-energy edge excitations.
If we simply act on a edge excitation with an
operator such as the lowest Landau level projected
density operator, the resulting state will
be in the lowest Landau level, but it will no
longer be annihilated by the Hamiltonian (\ref{halrezham}).
Hence we must project the resulting state into
the space of edge excitations annihilated
by (\ref{halrezham}). This projection cannot be performed
without a knowledge of the inner products of states,
so we are stuck again.

Ordinarily, this would not worry us too much
because the commutator algebra of the
resulting projected operators would be
more or less canonical and easily guessed.
However, in the case of the Haldane-Rezayi state,
the $SU(2)$ spin symmetry must be realized in
an unusual way because the edge theory
contains two real, i.e. Majorana, fermions,
say ${\psi_1}(x)$ and ${\psi_2}(x)$.
Their Lagrangian is invariant under the $O(2)$ rotations
${\psi_i}' = {O_{ij}}{\psi_j}$. There is
no local\footnote{i.e. so that ${\psi_i}'(x)$ depends only
${\psi_j}(x)$ and not on ${\psi_j}(x')$ for $x'\neq x$}
$SU(2)$ transformation law which preserves the reality property
of the Majorana spinors. The simplest way of having an $SU(2)$
doublet of fermions is to have two complex, i.e. Dirac,
spinors, $\chi_i$, which transform as
${\chi_i}' = {U_{ij}}{\chi_j}$. However, since a single Dirac
spinor is composed of two Majorana spinors, such a theory will have
too many states at each energy level.
Since the $SU(2)$ symmetry cannot be realized in the
standard way, the algebra of the spin-densities
in the $c=1$ theory can be and -- as we will see momentarily --
is anomalous.

First, however, we must answer a more basic question: if, as
we have conjectured, the $c=1$ Dirac theory is the
correct edge theory, where is the $SU(2)$ symmetry?
The answer is that the symmetry is hidden and non-local.
The Dirac theory has the Hamiltonian
\be
H = {\sum_{k}}\, vk\,{\psi_{k}^\dagger}{\psi_{k}}
\ee
where $\psi$ is a complex chiral fermion or,
equivalently, two real fermions, $\psi_1$, $\psi_2$,
with $\psi = {\psi_1} + i{\psi_2}$, and $v$ is the (non-universal)
velocity of the neutral fermions.
The generators of the $SU(2)$ symmetry are:
\bea
{S^z} = {\sum_{k}}\,{\psi_k^\dagger}
{\psi_k}\br
{S^+} = {\sum_{k>0}}\, {\psi_k^\dagger}{\psi_{-k}^\dagger}\br
{S^-} = {\sum_{k>0}}\, {\psi_{-k}}{\psi_{k}}
\eea
These generators commute with the Hamiltonian and, thus,
generate a global $SU(2)$ symmetry of the theory. These symmetry
generators were constructed in \cite{ludwig}.

In mapping the neutral sector of the edge theory onto
the $c=1$ Dirac fermion, we associate the up-spin
neutral fermions with the particles, created by ${\psi^\dagger_k}$
with $k>0$. The down-spin neutral fermions are associated with the
anti-particles, created by $\psi_{-k}$ ($k>0$). The
$SU(2)$ symmetry of the theory rotates up-spin fermions
into down-spin fermions, i.e. it mixes particles and anti-particles.
As a result, the transformation law is not local.

We can reformulate the Dirac theory in such
a way that this $SU(2)$ symmetry is more transparent.
We map ${\psi_k} \rightarrow {\chi_{1k}}$ and
${\psi_{-k}^\dagger} \rightarrow {\chi_{2k}}$
where $k>0$. The fields $\chi_\alpha$ form an
$SU(2)$ doublet with Hamiltonian
\be
H = {\sum_{k>0}}\, vk\,{\chi_{ik}^\dagger}{\chi_{ik}^{}}
\ee
and symmetry generators
\be
{S^a} = \sum_{k>0}\,{\chi_{\alpha k}^\dagger}\,
{\sigma^a_{\alpha\beta}}\,{\chi_{\beta k}^{}}
\label{symmgen}
\ee
As a result of the restriction to $k>0$, they are `half'
of an $SU(2)$ doublet of Dirac fermions. The
$k<0$ part of the theory has been discarded.
It then follows that there is no local $SU(2)$ Kac-Moody
algebra. If we introduce local spin densities,
${S^a}(x)$ and their Fourier transforms,
\be
{S^a_q} = \sum_{k>0}\,{\chi_{\alpha k+q}^\dagger}\,
{\sigma^a_{\alpha\beta}}\,{\chi_{\beta k}^{}}
\ee
we find that their commutators do not close
because the sums over $k$ are restricted to
$k>0$. In particular, their commutators are not local,
i.e. $[{S^a}(x),{S^b}(x')] \propto 1/(x-x')$
rather than $[{S^a}(x),{S^b}(x')] \propto \delta(x-x')$.

This might appear to be a death blow to the $c=1$
theory of the neutral sector. In the underlying
quantum mechanics of electrons, these commutators
are local, i.e. proportional to $\delta$-functions,
so we would expect that in the low-energy theory
they would be, at worst, $\delta$-functions smeared
out at the scale of the cutoff.
However, this argument is a bit too quick. The cutoff in this theory
is $O({V_1})$ (see (\ref{halrezham})) meaning that
our edge theory is an effective field theory
for energies less than $O({V_1})$. However, unlike in a Euclidean
or relativistic theory, this energy scale does not
imply a length scale. While the theory must be
local in time (again, modulo non-localities at scales
smaller than the cutoff), it does not necessarily
have commutators which are local in space.

But do the spin-densities, projected into the low-energy
subspace actually have such a non-local algebra? If not,
the $c=1$ theory must be ruled out. If so -- and, as we argued
above this would not contradict any fundamental
principle which is dear to our hearts -- then the $c=1$
theory is a viable candidate to describe the neutral sector
of the edge of the Haldane-Rezayi state. Consider
the following state, where ${{\cal P}_H}$ is
the projection operator into the zero-energy subspace
of (\ref{halrezham}):
\bea
{{\cal P}_H}\, {S^+}(w)\, {{\cal P}_H} \,\,\cdot\,\,
{\Psi_0} = \br {{\cal P}_H}\,
{\cal A}\left( {e^{ (2w{z_1}-{|w|^2}-{|{z_1}|^2})/4{{\ell}_0^2} }}
\frac{{u_1}{u_2}}{(w-{z_2})^2}\,\,
\frac{{u_3}{v_4}-{v_3}{u_4}}{({z_3} - {z_4})^2}\,\,\ldots\right)
{\prod_{i>1}}{\left(w-{z_i}\right)^2}\,
{\prod_{k>l>1}}{\left({z_k}-{z_l}\right)^2}\,\,
{e^{-\frac{1}{4{\ell}_0^2} {\sum_{i>1}} |z_i|^2 }}
\eea
which results from acting on the ground state with
the local projected ${S^+}(w)$ operator. It is quite plausible
that the right-hand-side vanishes upon projection. This would
agree with the $c=1$ theory, where ${S^+}(x)|0\rangle =
{\sum_{k>0,q}}\,{e^{iqx}}\,{\chi_{\alpha k}^\dagger}\,
{\sigma^+_{\alpha\beta}}\,{\chi_{\beta k}^{}}|0\rangle=0$
because ${\chi_{\beta k}^{}}|0\rangle=0$ for $k>0$. For
a doublet of Dirac fermions (and presumably for any other
theory with a local $SU(2)$ transformation law),
on the other hand, the $k<0$ modes will give a non-zero contribution.
Furthermore, suppose we act with this operator on a state
with $1$ neutral fermion:
\bea
{{\cal P}_H}\, {S^+}(w)\, {{\cal P}_H} \,\,\cdot\,\,
{\cal A}\left({z_1^k}{v_1}\,
\frac{{u_2}{v_3}-{v_2}{u_3}}{({z_2} - {z_3})^2}\,\,\ldots\right)
\prod{({z_i}-{z_j})^2} = \br
{{\cal P}_H}{\cal A}\left(
{e^{ (2w{z_1}-{|w|^2}-{|{z_1}|^2})/4{{\ell}_0^2} }}\,{w^k}{u_1}\,
\frac{{u_2}{v_3}-{v_2}{u_3}}{({z_2} - {z_3})^2}\,\,\ldots\right)
{\prod_{i>1}}{\left(w-{z_i}\right)^2}\,
{\prod_{k>l>1}}{\left({z_k}-{z_l}\right)^2}\,\,
{e^{-\frac{1}{4{\ell}_0^2} {\sum_{i>1}} |z_i|^2 }}\br
+\,\, {\rm \em terms\,\, in\,\, which\,\, the\,\, spin\,\, acts\,\,
on\,\, paired\,\, electrons} 
\eea
If this is non-vanishing, it is plausibly equal to
(the ${a_j}$ are some, possibly $w$-dependent, coefficients):
\be
{\cal A}\left(
{\sum_j}{a_j}{z_1^j}\,\,{w^k}{u_1}\,
\frac{{u_2}{v_3}-{v_2}{u_3}}{({z_2} - {z_3})^2}\,\,\ldots\right)
{\prod_{i>j}}{({z_i}-{z_j})^2}\,\,
{e^{-\frac{1}{4{\ell}_0^2} {\sum_{i}} |z_i|^2 }}
\ee
If so, then the up-spin electron (and its concomitant
neutral fermion) is no longer
localized at $w$ because of the large powers of $z_1$
from the Jastrow factor. In such a case, however,
when we act with another local spin operator, ${S^-}(w')$,
the commutator, which receives non-vanishing contributions
only when the two spin operators act on the same electron,
need not vanish for $w\neq w'$ (or, rather,
need not decay as a Gaussian in $w-w'$).

Even if our hypothesis is incorrect, and the Haldane-Rezayi
edge theory is some other theory, it is difficult
to see how the $SU(2)$ symmetry could be local.
There are simply `too few' single fermion states, by
a factor of two, to allow for a local $SU(2)$ symmetry.
This is quite clear from the formulation as a truncated Dirac
doublet. If we were to take an inner product different from
the inner product of the $c=1$ theory, this would
not help matters since it could not increase the size
of the Hilbert space. Could it be that we have simply
chosen the wrong symmetry generators? This is unlikely
since the symmetry generators (\ref{symmgen}) have
the desired action: they rotate the spins of the fermions.
In principle, there is one other possibility.
If there were low-energy excitations in the bulk
withh anomalous total derivative terms in
their $SU(2)$ algebra, these terms could cancel the
anomalous terms at the edge. However, there is no
trace of such excitations among the states annihilated
by (\ref{halrezham}).

\section{Experimental Consequences.}

If our hypothesis is correct and the edge theory of
the Haldane-Rezayi state is the $c=1+1$ conformal field
theory, there are measurable consequences which could
elucidate the nature of the $\nu=\frac{5}{2}$ plateau.
The electron annihilation operator is $\psi {e^{-i\sqrt{2}\phi}}$,
so the coupling to a Fermi liquid lead will be 
$\psi {e^{-i\sqrt{2}\phi}}\,{\Psi_{\rm lead}}$.
This is a dimension $2$ operator, so the tunneling conductance, $G_t$,
through a point contact between a Fermi liquid lead and the edge of the
Haldane-Rezayi state is
\be
{G_t} \sim {T^2}
\label{tuncond}
\ee
See \ct{wenedge,kanefisher} to compare \rf{tuncond} with the
corresponding expression for a Laughlin state. 
If the voltage $V \gg T$, then $I \sim {V^3}$. For tunneling between
two Haldane-Rezayi droplets, ${G_t} \sim {T^4}$ for
$T\gg V$ and $I \sim {V^5}$ for $T \ll V$. The tunneling
of quasiparticles from one edge of a Haldane-Rezayi droplet
to another through the bulk is presumably dominated
by the tunneling of half-flux quantum quasiparticles,
which are created by $\mu {e^{-i\phi/2\sqrt{2}}}$
where $\mu$ is the Dirac theory twist field. The
resulting tunneling conductance between the two edges
is ${G_t} \sim {T^{-5/4}}$ at high temperatures; at low
temperature it is $\frac{1}{2}\,\frac{e^2}{h}$ with
corrections determined by the perturbations of a
strong-coupling fixed point.

One thing which is, perhaps, surprising about these predictions
is that they are precisely the same as would be expected
for the $(3,3,1)$ state and for a simple reason:
the edge theories are almost the same. According to \cite{milovan},
the neutral sector of the $(3,3,1)$ state is a $c=1$
Dirac fermion. The only difference with the edge theory
of the Haldane-Rezayi state is that the twisted and
untwisted sectors are exchanged, but this does not
affect the dimensions of the scaling operators
which determine the above power laws. Hence, the
Haldane-Rezayi and $(3,3,1)$ states cannot
be distinguished from simple tunneling experiments at the edge.
However, these states are definitely not in the
same universality class. Their bulk excitations have
different topological properties, as may be seen from
the ground state degeneracy on the torus. In an
Aharonov-Bohm experiment with two half-flux quantum
quasiholes, the phase resulting when one winds around another is
$3\pi i/4$ in the Haldane-Rezayi state (from (\ref{hrsmu}))
but $-\pi i/8$ in the $(3,3,1)$ state (and $0$ in the
Pfaffian state). In experiments with more than two quasiholes,
the full structure of the non-Abelian statistics of the
Haldane-Rezayi state comes into play and, again,
Aharonov-Bohm experiments can resolve it from
the $(3,3,1)$ and other candidate quantum Hall states.

\begin{acknowledgements}
This work would not have been possible without
the help, advice, and encouragement,
given to us by a number of people, particularly L. Balents,
M.P.A. Fisher, K. Gawedzki, F.D.M. Haldane, B.I. Halperin, A.W.W. Ludwig,
M. Moriconi, J. Polchinski, V. Sadov, T. Spencer, A.M. Tikofsky,
P.B. Wiegmann, F. Wilczek, and A.B. Zamolodchikov.
\end{acknowledgements}


\begin{thebibliography} {99}
\bibitem{ludwig} While we were completing this paper, a preprint,
hep-th/9612172, by S. Guruswamy and A.W.W. Ludwig appeared
in which the relationship between the $c=1$ and $c=-2$
theories was discussed and the hidden symmetry of the
$c=1$ theory was constructed.
\end{thebibliography}
\end{document}